\documentclass[12pt]{article}
\usepackage{amsmath,amssymb,bm}
\usepackage{cite}
\title{Hawking Radiation from Kerr-Newman Black Hole and Tunneling Mechanism}
\date{}
\author{\bf Koichiro Umetsu\footnote{E-mail: umetsu@phys.cst.nihon-u.ac.jp}}

\makeatletter
 
  \@addtoreset{equation}{section}
 \makeatother
 
\begin{document}
\maketitle

\centerline{\it Institute of Quantum Science, College of Science and Technology}
\centerline{\it Nihon University, Chiyoda-ku, Tokyo 101-8308, Japan}

\vskip 0.5 truecm

\abstract{We present the derivation of Hawking radiation by using the tunneling mechanism 
in a rotating and charged black hole background.
We show that the 4-dimensional Kerr-Newman metric, which has a spherically nonsymmetric geometry, becomes an effectively 2-dimensional spherically symmetric metric by using the technique of the dimensional reduction near the horizon.
We can thus readily apply the tunneling mechanism to the nonspherical Kerr and Kerr-Newman metric.
}
\section{Introduction}

Hawking radiation is one of the most interesting effects 
which arise from the combination of quantum mechanics and general relativity.
Hawking's original derivation, which calculates the Bogoliubov coefficients between the in- and out-states for a body collapsing to form a black hole, is very direct and physical \cite{Haw01}.
It is well-known that the characteristic spectrum obtained from the original derivation agrees with the blackbody spectrum with the Hawking temperature, if we ignore the backscattering of particles falling into the horizon.

After Hawking's original derivation, several derivations of Hawking radiation have been suggested
\cite{Chr01,Haw02,Ban01,Str01,SP, Ter01, Rob01}.
In particular, the derivation by the tunneling mechanism provides one of intuitive derivations 
\cite{SP, T01, T02, T03, T04, T05, T06, T07, T08, T09, T10, T11, T12, T13, T14, T15, T16, T17, T18, T19, T20,
T21, T22, T23, T24, T25, T26, T27, BM01, T28, T29}.
The essential idea of the tunneling mechanism is that a particle-antiparticle pair is formed close to the horizon.
The ingoing mode is trapped inside the horizon while the outgoing mode can quantum mechanically tunnel through the horizon
and it is observed at infinity as the Hawking flux. 
As described above, this derivation is intuitively understandable.
However, in the literature \cite{SP, T01, T02, T03, T04, T05, T06, T07, T08, T09, T10, T11, T12, T13, T14, T15, T16, T17, T18, T19, T20, T21, T22, T23, T24, T25, T26, T27}, the analysis is confined to the derivation of the Hawking temperature
only by comparing the tunneling probability of an outgoing particle with the Boltzmann factor.
There exists no discussion of the spectrum.
Therefore, there remains the possibility that the black hole is not the black but merely the thermal body.
This problem was pointed out by Banerjee and Majhi \cite{BM01}.
They directly showed how to reproduce the black body spectrum with the Hawking temperature from the expectation value of number operator
by using the properties of the tunneling mechanism.
Thus the derivation from the tunneling mechanism became more satisfactory.

Their result is valid for black holes with spherically symmetric geometry such as Schwarzschild or Reissner-Nordstr\"om black holes in the 4-dimensional theory.
However, 4-dimensinal black holes may have not only a mass but also a charge and the angular momentum
according to the black hole uniqueness theorem (the no hair theorem) \cite{Car01,Ruf01}.
In 4-dimensions, the Kerr-Newman black hole, which has both the charge and angular momentum, is the most general black hole
and its geometry becomes spherically nonsymmetric because of its own rotation.
There exist several previous works for a rotating black hole in the tunneling method (see for example \cite{T08, T21, T28}), 
but they are mathematically very involved.

We would like to extend the simple derivation of Banerjee and Majhi by the tunneling mechanism to the case of the Kerr-Newman black hole.
It is known that the 4-dimensional Kerr-Newman metric effectively becomes a 2-dimensional spherically symmetric metric by using the technique of the dimensional reduction near the horizon \cite{iso01,ume01}.
The essential idea is as follows: We consider the action for a scalar field. 
We can then ignore the mass, potential and interaction terms in the action because the kinetic term dominates in the high-energy theory near the horizon.
By expanding the scalar field in terms of the spherical harmonics and using properties at horizon,
we find that the integrand in the action dose not depend on angular variables.
Thus we find that the 4-dimensional action with the Kerr-Newman metric effectively becomes a 2-dimensional action with the spherically symmetric metric.
This technique was often used in the derivation of Hawking radiation from anomalies \cite{Rob01,iso01,ume01,ano01,ano02,ano03,ano04,ano05,ano06,ano07}.

It is possible to use the same technique in the tunneling mechanism 
since the tunneling effect is also the quantum effect arising within the Planck length near the horizon region.
By this procedure, the metric for the Kerr-Newman black hole becomes an effectively 2-dimensional spherically symmetric metric,
and we can use the approach of Banerjee and Majhi which is valid for black holes with spherically symmetric geometry. 
We can thus derive the black body spectrum and Hawking flux for the Kerr-Newman black hole in the tunneling mechanism.

The content of the present paper is as follows.
In section 2, we show that the 4-dimensional Kerr-Newman metric effectively becomes a 2-dimensional spherically symmetric metric
by using the technique of the dimensional reduction near the horizon.
In section 3, we discuss the relation between states inside and outside the black hole in connection with the tunneling effect in the induced metric.
In section 4, we show how to derive the blackbody spectrum and the Hawking flux for the Kerr-Newman black hole
by following Banerjee and Majhi's approach.
Section 5 is devoted to conclusions and discussions.
In Appendix A, we show how to obtain the Kruskal-like coordinates for the Kerr-Newman black holes.

\section{Dimensional Reduction near the Horizon}

In this section, 
we show that the 4-dimensional Kerr-Newman metric becomes a 2-dimensional spherically symmetric metric 
by using technique of the dimensional reduction near the horizon \cite{iso01}.

For a rotating and charged black hole, the metric is given by the Kerr-Newman metric
\begin{align}
ds^2=&-\frac{\Delta -a^2 \sin^2\theta}{\Sigma}dt^2-\frac{2a\sin^2\theta}{\Sigma}(r^2+a^2-\Delta)dtd\varphi \notag\\
&-\frac{a^2\Delta \sin^2\theta-(r^2+a^2)^2}{\Sigma}\sin^2\theta d\varphi^2+\frac{\Sigma}{\Delta}dr^2+\Sigma d\theta^2,
\label{KN01}
\end{align}
where notations are respectively defined by
\begin{align}
a&\equiv \frac{J}{M},
\label{KN02}\\
\Sigma&\equiv r^2+a^2\cos^2 \theta ,
\label{KN03}\\
\Delta&\equiv r^2-2Mr+a^2+Q^2=(r-r_+)(r-r_-).
\label{KN04}
\end{align}
Also, $M$, $J$, $Q$ and $r_{+(-)}$ are respectively a mass, an angular momentum, an electrical charge and the outer (inner) horizon of the Kerr-Newman black hole
\begin{align}
r_\pm=M\pm \sqrt{M^2-a^2-Q^2}.
\label{KN05}
\end{align}
It follows from the expression (\ref{KN01}) that the Kerr-Newman metric is spherically non-symmetric geometry.

For simplicity, we consider the 4-dimensional action for a complex scalar field
\begin{align}
S=\int d^4x \sqrt{-g}g^{\mu\nu}( \partial_\mu +ieV_\mu ) \phi^* ( \partial_\nu -ieV_\nu )\phi +S_{{\rm int}},
\label{KNact01}
\end{align}
where the first term is the kinetic term and the second term $S_{{\rm int}}$ represents the mass, potential and interaction terms.
The gauge field $V_\mu$ associated with the Coulomb potential of the black hole, is given by
\begin{align}
( V_{\mu} )=\left( -\frac{Qr}{r^2+a^2}, 0, 0, 0\right).
\label{gauge01}
\end{align}

By substituting both (\ref{KN01}) and (\ref{gauge01}) to (\ref{KNact01}), we obtain
\begin{align}
S=\int dt dr d\theta d\varphi \sin \theta \phi^* 
\Bigg[ & \left( \frac{(r^2+a^2)^2}{\Delta}-a^2\sin^2 \theta \right)  \left( \partial_t+\frac{ieQr}{r^2+a^2} \right) ^2\notag\\
&+2ia \left( \frac{r^2+a^2}{\Delta} -1 \right) \left( \partial_t+\frac{ieQr}{r^2+a^2} \right) \hat{L}_z
-\partial_r\Delta \partial_r +\hat{\bm L}^2 -\frac{a^2}{\Delta}\hat{L}^2_z
\Bigg]\phi+S_{{\rm int}},
\label{kerract2}
\end{align}
where we used
\begin{align}
\hat{\bm L}^2&=-\frac{1}{\sin \theta}\partial_\theta \sin \theta \partial_\theta-\frac{1}{\sin^2 \theta}\partial_\varphi^2,\\
\hat{L}_z&=-i\partial_\varphi.
\end{align}
By performing the partial wave decomposition of $\phi$ in terms of the spherical harmonics
\begin{align}
\phi=\sum_{l,m}\phi_{lm}(t,r)Y_{lm}(\theta,\varphi),
\label{kyu1}
\end{align}
we obtain
\begin{align}
S=&\int dt dr d\theta d\varphi \sin\theta \sum_{l',m'}\phi_{l'm'}^* Y_{l'm'}^*\notag\\
&\times
\Bigg[ \frac{(r^2+a^2)^2}{\Delta}\left( \partial_t+\frac{ieQr}{r^2+a^2} \right)^2-a^2\sin^2 \theta \left( \partial_t+\frac{ieQr}{r^2+a^2} \right)^2 
+2ima\frac{r^2+a^2}{\Delta}\left( \partial_t+\frac{ieQr}{r^2+a^2} \right)\notag\\
&\qquad -2ima\left( \partial_t+\frac{ieQr}{r^2+a^2} \right)
-\partial_r\Delta \partial_r +l(l+1) -\frac{m^2a^2}{\Delta}
\Bigg]\sum_{l,m}\phi_{lm}Y_{lm}+S_{{\rm int}},
\label{kerract4}
\end{align}
where we used eigenvalue equations for $\hat{\bm L}^2$ and $\hat{L}_z$
\begin{align}
\hat{\bm L}^2Y_{lm}&=l(l+1)Y_{lm},
\label{kaku1}\\
\hat{L}_zY_{lm}&=mY_{lm}.
\label{kaku2}
\end{align}
Here $l$ is the azimuthal quantum number and $m$ is the magnetic quantum number.
Now, we transform the radial coordinate $r$ into the tortoise coordinate $r_*$ defined by
\begin{align}
\frac{dr_*}{dr}=\frac{r^2+a^2}{\Delta}\equiv\frac{1}{F(r)}.
\label{fdef}
\end{align}
After this transformation, the action (\ref{kerract4}) is written by
\begin{align}
S=&\int dt dr_* d\theta d\varphi \sin\theta \sum_{l'm'}\phi_{l'm'}^*Y_{l'm'}^*\notag\\
&\times \Bigg[ (r^2+a^2)\left( \partial_t+\frac{ieQr}{r^2+a^2} \right)^2 
-F(r)a^2\sin^2\theta \left( \partial_t+\frac{ieQr}{r^2+a^2} \right)^2 
+2ima\left( \partial_t+\frac{ieQr}{r^2+a^2} \right)\notag\\
&
\qquad-F(r)2ima \left( \partial_t+\frac{ieQr}{r^2+a^2} \right)
-\partial_{r_*}(r^2+a^2)\partial_{r_*}+F(r)l(l+1)-\frac{m^2a^2}{r^2+a^2}
\Bigg]\sum_{l,m}\phi_{lm}Y_{lm}\notag\\
&+S_{{\rm int}}.
\label{kerract5}
\end{align}

Here we consider this action in the region near the horizon.
Since $F(r_+)=0$ at $r \to r_+$,  we only retain dominant terms in (\ref{kerract5}).
We thus obtain
\begin{align}
S=\int dt dr_* d\theta d\varphi \sin\theta \sum_{l'm'}\phi_{l'm'}^*Y_{l'm'}^*
\Bigg[ &(r^2+a^2)\left( \partial_t+\frac{ieQr}{r^2+a^2} \right)^2 +2ima\left( \partial_t+\frac{ieQr}{r^2+a^2} \right)\notag\\
&-\partial_{r_*}(r^2+a^2)\partial_{r_*}-\frac{m^2a^2}{r^2+a^2}
\Bigg]\sum_{l,m}\phi_{lm}Y_{lm},
\label{kerract6}
\end{align}
where we ignored $S_{{\rm int}}$ by using $F(r_+)=0$ at $r \to r_+$. 
Because the theory becomes the high-energy theory near the horizon and the kinetic term dominates, 
we can ignore all the terms in $S_{{\rm int}}$.
After this analysis, we return to the expression written in terms of $r$, and we obtain
\begin{align}
S=-\sum_{l,m}\int dt dr(r^2+a^2)\phi_{lm}^*
\left[ -\frac{r^2+a^2}{\Delta}\left( \partial_t +\frac{ieQr}{r^2+a^2} +\frac{ima}{r^2+a^2}\right)^2+\partial_r \frac{\Delta}{r^2+a^2} \partial_r\right]\phi_{lm},
\label{kerract9}
\end{align}
where we used the orthonormal condition for the spherical harmonics
\begin{align}
\int d\theta d\varphi \sin\theta Y_{l'm'}^*Y_{lm}=\delta_{l',l}\delta_{m',m}.
\end{align}

From (\ref{kerract9}), we find that $\phi_{lm}$ can be considered as a (1+1)-dimensional complex scalar field in the backgrounds of 
the dilaton $\Phi$, metric $g_{\mu\nu}$ and two $U(1)$ gauge fields $V_\mu$, $U_\mu$
\begin{align}
&\Phi=r^2+a^2,\\
&g_{tt}=-F(r),\ 
g_{rr}=\frac{1}{F(r)},\ 
g_{rt}=0,
\label{kerr001}
\\
&V_t =-\frac{Qr}{r^2+a^2}, \ V_r =0,\\
&U_t =-\frac{a}{r^2+a^2}, \ U_r =0.
\end{align}
There are two $U(1)$ gauge fields:
One is the original gauge field as in (\ref{gauge01})
while the other is the induced gauge field associated with the isometry along the $\varphi$ direction.
The induced $U(1)$ charge of the 2-dimensional field $\phi_{lm}$ is given by $m$.
Then the gauge potential $A_t$ is a sum of these two fields,
\begin{align}
A_t\equiv eV_t+mU_t=-\frac{eQr}{r^2+a^2}-\frac{ma}{r^2+a^2}, \quad A_r=0.
\label{gauge02}
\end{align}

From (\ref{kerr001}), we find that the 4-dimensional spherically non-symmetric Kerr-Newman metric (\ref{KN01}) behaves as
a 2-dimensional spherically symmetric metric in the region near the horizon
\begin{align}
ds^2=-F(r)dt^2+\frac{1}{F(r)}dr^2.
\label{kerr5}
\end{align}

For confirmation, we show how to derive the surface gravity on the horizon of the Kerr-Newman black hole from $F(r)$ defined by (\ref{fdef}).
Actually by calculating the surface gravity, we can obtain
\begin{align}
K_\pm \equiv \frac{1}{2} F'(r)\Bigg |_{r=r_\pm}=\frac{r_\pm-r_\mp}{2(r_\pm^2+a^2)},
\end{align}
where $\{'\}$ represents differentiation with respect to $r$.
This result agrees with the well-known surface gravity on the horizon of the Kerr-Newman black hole. 


\section{Tunneling Mechanism}

In this section, we discuss the connection between states inside and outside the black hole to analyze the tunneling effect in the induced metric.
We consider the Klein-Gordon equation near the horizon. 
In previous section, 
we showed that we can regard the 4-dimensional Kerr metric as the 2-dimensional spherically symmetric metric in the region near the horizon.
As already stated, 
since the kinetic term dominates in the high-energy theory near the horizon, we can ignore the mass, potential and interaction terms.
We obtain the Klein-Gordon equation with the gauge field from the action (\ref{kerract9})
\begin{align}
\left[ \frac{1}{F(r)}\left( \partial_t -iA_t\right)^2-F(r)\partial_r^2 - F'(r) \partial_r \right] \phi=0,
\label{KGeq01}
\end{align}
where $A_t$ is defined in (\ref{gauge02}).
Of course, this equation can be obtained from the general Klein-Gordon equation for a free particle with the gauge field in 2-dimensional space-time
\begin{align}
g^{\mu\nu}(\nabla_\mu-iA_\mu)(\nabla_\nu-iA_\nu)\phi=0
\end{align}
in a manner similar to \cite{BM01}.
Taking the standard WKB ansatz
\begin{align}
\phi(r,t)=e^{\frac{i}{\hbar}S(r,t)},
\end{align}
and substituting the expansion of $S(r,t)$
\begin{align}
S(r,t)=S_0(r,t)+\sum^\infty_{i=1}\hbar^i S_i(r,t)
\end{align}
in (\ref{KGeq01}), we obtain, in the semiclassical limit (i.e. $\hbar\to 0$),
\begin{align}
\partial_t S_0 (r,t)=\pm F(r) \partial_r S_0 (r,t).
\label{KGeq02}
\end{align}
We find that terms including the gauge field vanished in the semiclassical limit.
This equation completely agrees with the equation in \cite{BM01} although the content of $F(r)$ is different from that used in \cite{BM01}.
From the Hamilton-Jacobi equation \cite{SP} and (\ref{KGeq02}), we can solve 
\begin{align}
S_0(r,t)=\left( \omega -e\Phi - m\Omega \right)(t\pm r_*)\equiv \omega'(t\pm r_*),
\label{ome'}
\end{align}
where $r_*$ is the tortoise coordinate defined by (\ref{fdef}), $\omega$ is the characteristic frequency,
$\Phi$ is the electric potential,
and $\Omega$ is the angular frequency on the horizon written as
\begin{align}
\Phi=\frac{Qr_+}{r_+^2+a^2},\quad \Omega=\frac{a}{r_+^2+a^2}.
\end{align}
Thus we obtain the semiclassical solution for the scalar field
\begin{align}
\phi(r,t)=\exp\left[ -\frac{i}{\hbar} \omega' (t \pm r_*) \right].
\label{phi01}
\end{align}

If one considers the metric at all regions for the charged and rotating black hole,
the exterior metric of the horizon is given by the 4-dimensional Kerr-Newman metric (\ref{KN01}).
However, we showed that the metric near the horizon can be regarded as the 2-dimensional spherically symmetric metric as in (\ref{kerr5}).
We thus use the metric (\ref{kerr5}) in the region near the horizon.
But we do not know the interior metric of the black hole.
However, if the interior metric is smoothly connected to the exterior metric through the horizon,
we may be able to identify the interior metric near the horizon to be the same as the exterior metric near the horizon.
We thus suppose that the interior metric near the horizon is given by (\ref{kerr5}).
Furthermore, since the tunneling effect is the quantum effect arising within the Plank length in the near horizon region,
we have to consider both the inside and outside regions which are very close to the horizon.

Here we use both the retarded time $u$ and the advanced time $v$ defined by
\begin{align}
u\equiv t-r_*,\qquad v\equiv t+r_*.
\label{uv}
\end{align}
We can then separate the scalar field (\ref{phi01}) into the ingoing (left handed) modes and outgoing (right handed) modes.
In the regions $r_+-\varepsilon < r <r_+$, and $r_+ \leq r <r_+-\varepsilon$, respectively, we express the field $\phi$ as
\begin{align}
& \phi^{R} _{\rm{in}}=\exp \left( -\frac{i}{\hbar} \omega' u_{\rm{in}}\right),\qquad
 \phi^{L}_{\rm{in}}=\exp \left( -\frac{i}{\hbar} \omega' v_{\rm{in}}\right),\qquad (r_+-\varepsilon < r <r_+),\\
& \phi^{R}_{\rm{out}}=\exp \left( -\frac{i}{\hbar} \omega' u_{\rm{out}}\right),\quad
\phi^{L} _{\rm{out}}=\exp \left( -\frac{i}{\hbar} \omega' v_{\rm{out}}\right), \quad (r_+ \leq r <r_++\varepsilon),
\end{align}
where $\varepsilon$ is an arbitrarily small constant, "R (L)" stand for the right (left) modes and "in (out)" stand for the inside (outside) of the black hole (Fig.1).

\begin{center}
\vspace{0.6cm}

\unitlength 0.1in
\begin{picture}( 29.3000, 23.2500)( 12.7000,-27.0500)
%
\special{pn 8}%
\special{sh 0.600}%
\special{pa 3200 2600}%
\special{pa 1270 2600}%
\special{pa 1270 380}%
\special{pa 3200 380}%
\special{pa 3200 2600}%
\special{ip}%
\put(17.4000,-15.8000){\makebox(0,0){{\bf BH}}}%
%
\special{pn 8}%
\special{pa 3792 1276}%
\special{pa 4150 1276}%
\special{fp}%
\special{sh 1}%
\special{pa 4150 1276}%
\special{pa 4084 1256}%
\special{pa 4098 1276}%
\special{pa 4084 1296}%
\special{pa 4150 1276}%
\special{fp}%
%
\special{pn 8}%
\special{pa 3670 1276}%
\special{pa 3330 1276}%
\special{fp}%
\special{sh 1}%
\special{pa 3330 1276}%
\special{pa 3398 1296}%
\special{pa 3384 1276}%
\special{pa 3398 1256}%
\special{pa 3330 1276}%
\special{fp}%
\put(37.3000,-15.6000){\makebox(0,0){outside}}%
\put(35.2000,-11.2000){\makebox(0,0){$\phi^{L} _{\rm{out}}$}}%
\put(39.4000,-11.2000){\makebox(0,0){$\phi^{R} _{\rm{out}}$}}%
%
\special{pn 20}%
\special{pa 3200 400}%
\special{pa 3200 2600}%
\special{fp}%
%
\special{pn 8}%
\special{pa 2730 1280}%
\special{pa 3088 1280}%
\special{fp}%
\special{sh 1}%
\special{pa 3088 1280}%
\special{pa 3022 1260}%
\special{pa 3036 1280}%
\special{pa 3022 1300}%
\special{pa 3088 1280}%
\special{fp}%
%
\special{pn 8}%
\special{pa 2600 1280}%
\special{pa 2260 1280}%
\special{fp}%
\special{sh 1}%
\special{pa 2260 1280}%
\special{pa 2328 1300}%
\special{pa 2314 1280}%
\special{pa 2328 1260}%
\special{pa 2260 1280}%
\special{fp}%
\put(26.7000,-15.7000){\makebox(0,0){inside}}%
\put(24.7000,-11.2000){\makebox(0,0){$\phi^{L} _{\rm{in}}$}}%
\put(29.0000,-11.3000){\makebox(0,0){$\phi^{R} _{\rm{in}}$}}%
%
\special{pn 8}%
\special{pa 2190 390}%
\special{pa 2190 2590}%
\special{dt 0.045}%
%
\special{pn 8}%
\special{pa 4200 390}%
\special{pa 4200 2590}%
\special{dt 0.045}%
%
\special{pn 8}%
\special{pa 3200 2190}%
\special{pa 4200 2190}%
\special{fp}%
\special{sh 1}%
\special{pa 4200 2190}%
\special{pa 4134 2170}%
\special{pa 4148 2190}%
\special{pa 4134 2210}%
\special{pa 4200 2190}%
\special{fp}%
\special{pa 4190 2190}%
\special{pa 3200 2190}%
\special{fp}%
\special{sh 1}%
\special{pa 3200 2190}%
\special{pa 3268 2210}%
\special{pa 3254 2190}%
\special{pa 3268 2170}%
\special{pa 3200 2190}%
\special{fp}%
\put(37.1000,-23.3000){\makebox(0,0){$\varepsilon$}}%
%
\special{pn 8}%
\special{pa 2176 2196}%
\special{pa 3176 2196}%
\special{fp}%
\special{sh 1}%
\special{pa 3176 2196}%
\special{pa 3108 2176}%
\special{pa 3122 2196}%
\special{pa 3108 2216}%
\special{pa 3176 2196}%
\special{fp}%
\special{pa 3166 2196}%
\special{pa 2176 2196}%
\special{fp}%
\special{sh 1}%
\special{pa 2176 2196}%
\special{pa 2242 2216}%
\special{pa 2228 2196}%
\special{pa 2242 2176}%
\special{pa 2176 2196}%
\special{fp}%
\put(26.8500,-23.3500){\makebox(0,0){$\varepsilon$}}%
\put(32.0000,-27.9000){\makebox(0,0){${\rm Horizon}$}}%
\put(41.9000,-27.9000){\makebox(0,0){$r_++\varepsilon$}}%
\put(21.8000,-27.9000){\makebox(0,0){$r_+-\varepsilon$}}%
\end{picture}%

\vspace{0.6cm}

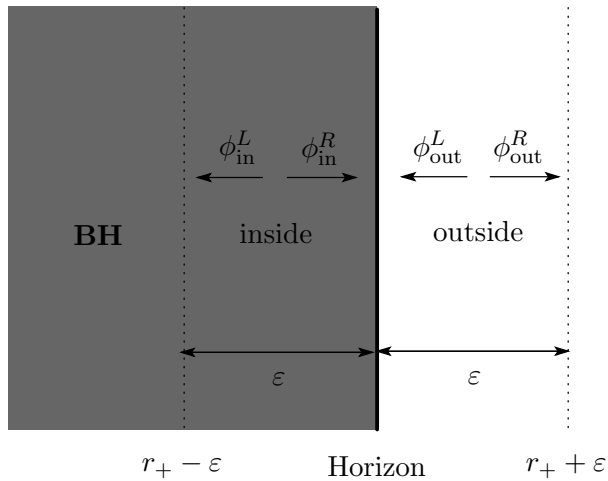
Fig.1 Intuitive picture of the scalar field near the horizon.

\vspace{0.6cm}
\end{center}

By performing an appropriate coordinate transformation (see appendix A), we can obtain the Kruskal-like coordinate variables \cite{Kru,Sze}
\begin{align}
&T_{\rm{in}}=\exp \left[ K_+ (r_*)_{\rm{in}}\right] \cosh \left( K_+ t_{\rm{in}} \right),\qquad \quad ~
R_{\rm{in}}=\exp \left[ K_+ (r_*)_{\rm{in}}\right] \sinh \left( K_+ t_{\rm{in}} \right),
\label{Kru01}\\
&T_{\rm{out}}=\exp \left[ K_+ (r_*)_{\rm{out}}\right] \sinh \left( K_+ t_{\rm{out}} \right),\qquad
R_{\rm{out}}=\exp \left[ K_+ (r_*)_{\rm{out}}\right] \cosh \left( K_+ t_{\rm{out}} \right),
\label{Kru02}
\end{align}
where $r_*$ is defined by (\ref{fdef}) and is explicitly given by
\begin{align}
r_*=r+\frac{1}{2K_+}\ln \frac{|r-r_+|}{r_+}+\frac{1}{2K_-}\ln\frac{|r-r_-|}{r_-},
\end{align}
and $K_{+(-)}$ is the surface gravity on the outer (inner) horizon.
Both of the relations (\ref{Kru01}) and (\ref{Kru02}) agree with the relations in \cite{BM01}.

In general, the Schwarzschild and Kerr-Newman metrics describe the behavior outside the black hole.
Consequently, the readers may wonder if these metrics defined by the Kruskal coordinates
can really describe the behavior inside the black hole.
In our case, however, we study the tunneling effect across the horizon
and thus we study the behavior in the very small regions near the horizon.
In such analysis, due to the reasons of continuity, it may not be unnatural to assume that
the Kruskal coordinates (3.12) can be used in both of outside and inside regions of near the horizon.

A set of coordinates (\ref{Kru01}) are connected with the other coordinates (\ref{Kru02}) by the relations,
\begin{align}
t_{\rm{in}}\to t_{\rm{out}}-i\frac{\pi}{2K_+},\qquad
(r_*)_{\rm{in}}\to (r_*)_{\rm{out}}+i\frac{\pi}{2K_+},
\end{align}
so that, with this mapping, $T_{\rm{in}} \to T_{\rm{out}}$ and $X_{\rm{in}} \to X_{\rm{out}}$.
Following the definition (\ref{uv}), we obtain the relations connecting the null coordinates defined inside and outside the horizon,
\begin{align}
u_{\rm{in}}&\equiv t_{\rm{in}} -(r_*)_{\rm{in}} \to u_{\rm{out}}-i\frac{\pi}{K_+},
\\
v_{\rm{in}}&\equiv t_{\rm{in}} +(r_*)_{\rm{in}}\to v_{\rm{out}}.
\label{v}
\end{align}
Under these transformations the inside and outside modes are connected by,
\begin{align}
\phi^{R}_{\rm{in}}&\equiv \exp \left( -\frac{i}{\hbar} \omega' u_{\rm{in}}\right) 
\to \exp\left( - \frac{\pi\omega'}{\hbar K_+} \right) \phi^{R}_{\rm{out}},
\label{phiR}\\
\phi^{L}_{\rm{in}}&\equiv\exp \left( -\frac{i}{\hbar} \omega' v_{\rm{in}}\right) \to \phi^{L}_{\rm{out}}.
\label{phiL}
\end{align}
As already discussed by Banerjee and Majhi in \cite{BM01}, the essential idea of the tunneling mechanism is that a particle-antiparticle pair is formed close to the horizon.
This pair creation may arise inside the black hole ( in the region close to the horizon ), 
since the space-time is locally flat.
The ingoing mode is trapped inside the horizon while the outgoing mode can quantum mechanically tunnel through the horizon.
It is then observed at infinity as the Hawking flux. 
We find that the effect of the ingoing mode inside the horizon do not appear outside the horizon as in (\ref{phiL})
since $v_{\rm{in}}$ changes to $v_{\rm{out}}$ without an extra term under the transformation connecting the null coordinates defined inside and outside the horizon as in (\ref{v}).
On the other hand, we find that the effect of the outgoing mode inside the horizon appear with a non-negligible probability  
by tunneling through the horizon quantum mechanically as in (\ref{phiR}).
This consideration agrees with the concept of tunneling mechanism.
Furthermore, we showed that we can treat the Kerr-Newman metric as a 2-dimensional spherically symmetric metric 
with a 2-dimensional effective gauge field just as in the case of
a simplest Schwarzschild metric in tunneling mechanism.


\section{Black body spectrum and Hawking flux}

In this section, we show how to derive the Hawking black body spectrum for a Kerr-Newman black hole by following the approach of Banerjee and Majhi's  \cite{BM01}.
First, we consider $n$ number of non-interacting virtual pairs that are created inside the black hole.
Then the physical state of the system is conventionally written as
\begin{align}
| \Psi \rangle =N\sum_n |n_{\rm{in}}^L \rangle \otimes |n_{\rm{in}}^R \rangle,
\label{sta}
\end{align}
where $|n_{\rm{in}}^{L(R)} \rangle$ is the number state of left (right) going modes inside the black hole and
$N$ is a normalization constant. From the transformations of both (\ref{phiR}) and (\ref{phiL}), we obtain
\begin{align}
|\Psi \rangle =N \sum_n e^{-\frac{\pi n \omega'}{\hbar K_+}} |n_{\rm{out}}^L \rangle \otimes |n_{\rm{out}}^R \rangle.
\label{psi1}
\end{align}
Here $N$ can be determined by using the normalization condition $\langle \Psi | \Psi \rangle=1$.
It is natural to determine the normalization constant $N$ for the state outside the black hole, 
because the observer exists outside the black hole.
Thus we obtain
\begin{align}
N=  \frac{1}{\left( \displaystyle\sum _{n} e^{-\frac{2\pi n \omega'}{\hbar K_+}} \right) ^{\frac{1}{2}}} .
\end{align}
For bosons ($n=0, 1, 2, \cdots $), $N_{(\rm{boson})}$ is calculated as
\begin{align}
N_{(\rm{boson})}=\left( 1 - e^{-\frac{2\pi \omega'}{\hbar K_+}} \right)^{\frac{1}{2}}.
\label{Nbos}
\end{align}
By substituting (\ref{Nbos}) into (\ref{psi1}), we obtain the normalized physical states of a system of bosons
\begin{align}
|\Psi \rangle_{(\rm{boson})}&=\left( 1- e^{-\frac{2\pi \omega'}{\hbar K_+}} \right)^{\frac{1}{2}}\sum_{n} e^{-\frac{\pi n \omega'}{\hbar K_+}} |n_{\rm{out}}^{L}\rangle \otimes |n_{\rm{out}}^{R}\rangle.
\end{align}
The density matrix operator of the system is given by
\begin{align}
\hat{\rho}_{(\rm{boson})}&\equiv |\Psi \rangle_{(\rm{boson})}\langle \Psi |_{(\rm{boson})}\notag\\
&=\left( 1- e^{-\frac{2\pi \omega'}{\hbar K_+}} \right)\sum_{n,m}
e^{-\frac{\pi \omega'}{\hbar K_+}(n+m)}
|n_{\rm{out}}^{L}\rangle \otimes |n_{\rm{out}}^{R}\rangle\langle m_{\rm{out}}^{R} | \otimes \langle m_{\rm{out}}^{L}|.
\end{align}
By tracing out the left going modes, we obtain the reduced density matrix for the right going modes, 
\begin{align}
\hat{\rho}_{(\rm{boson})}^{(R)}
&={\rm Tr} \left( \hat{\rho}_{(\rm{boson})}^{(R)} \right)\\
&=\sum_{n}\langle n_{\rm{out}}^{L} | \hat{\rho}_{(\rm{boson})} |n_{\rm{out}}^{L}\rangle\\
&=\left( 1- e^{-\frac{2\pi \omega'}{\hbar K_+}} \right)\sum_{n}e^{-\frac{2\pi n \omega'}{\hbar K_+}}  |n_{\rm{out}}^{R}\rangle\langle n_{\rm{out}}^{R} |.
\label{red}
\end{align}
Then, the expectation value of the number operator $\hat{n}$ is given by
\begin{align}
\langle n \rangle_{{\rm boson}} &= {\rm Tr} \left( \hat{n} \hat{\rho}_{(\rm{boson})}^{(R)} \right)\\
&=\frac{1}{e^{\frac{2\pi \omega'}{\hbar K_+}} -1}\\
&=\frac{1}{e^{\beta \left( \omega -e \Phi - m \Omega \right)} -1}
\label{num3}
\end{align}
where in the last line we used the definition (\ref{ome'}) and we identify the Hawking temperature $T_H$ by
\begin{align}
\beta \equiv \frac{1}{T_H} \equiv \frac{2\pi}{\hbar K_+}.
\end{align}
This result corresponds to the black body spectrum with the Hawking temperature
and agrees with previous works in the Kerr-Newman black hole background \cite{Haw01}.

Moreover, the Hawking flux $F_H$ can be derived by integrating the sum of the distribution function for a particle with a quantum number ($e$, $m$) and its antiparticle with ($-e$, $-m$) over all $\omega$'s
\begin{align}
F_H&= \int^\infty _0 \frac{d\omega}{2\pi} \omega \left[ \frac{1}{e^{\beta(\omega-e\Phi-m\Omega)}-1}
+\frac{1}{e^{\beta(\omega+e\Phi +m\Omega)}-1} \right]\\
&=\frac{\pi}{6\beta^2}-\frac{1}{4\pi} (e\Phi+m\Omega)^2,
\end{align}
where we ignored the superradiance. However, as is well-known, boson fields display the superradiance provided that they have 
frequency in a certain range whereas fermion fields do not \cite{sup01,sup02,sup03}.
In the fermionic case also, one can derive the Hawking flux following the above derivation. It is then shown that the result agrees with the previous result \cite{iso01}.

Before closing, we discuss the black hole entropy.
Since a particle emitted by the black hole has the Hawking temperature, it is natural to consider that the black hole itself has the same temperature.
Thus we can obtain the black hole entropy $dS_{{\rm BH}}$ from a thermodynamic consideration
\begin{align}
dS_{{\rm BH}}=\left( \frac{dM}{T_H} \right).
\label{ent01}
\end{align}
By integrating Eq. (\ref{ent01}), the entropy agrees with the Bekenstein-Hawking entropy
\begin{align}
S_{{\rm BH}}=\frac{A_{{\rm BH}}}{4\hbar},
\label{ent02}
\end{align}
where $A_{{\rm BH}}$ is the surface area of the black hole
\begin{align}
A_{{\rm BH}} =4\pi (r_+^2+a^2).
\end{align}
We here mention that this entropy (\ref{ent02}) is not derived by counting the number of quantum states
associated with the black hole since the black hole is characterized solely by the classical solution of the metric in our analysis.

\section{Conclusion and Discussion}

We have shown how to derive the blackbody spectrum of the Hawking radiation and its flux for the Kerr-Newman black hole by combining the Banerjee and Majhi's tunneling approach with the technique of dimensional reduction.

In comparison with previous works, our derivation has several advantages.
First, Hawking's original derivation assumes that the self interaction of the matter field can be ignored \cite{Haw01}.
Also, the similar limitation appears in the recent derivations of Hawking radiation from anomalies \cite{ano06,ano07}.
For example, the derivation of Hawking radiation presented by Banerjee and Kulkarni,
which simplified and clarified the original derivation from anomalies using step functions \cite{Rob01} 
by making explicit a connection between the region near the horizon and the region away from the horizon, 
still assumes that the 4-dimensional theory away from the horizon can be treated as an effective 2-dimensional conformal field theory. 
They also used physically valid boundary conditions such as the Unruh vacuum.
However, they cannot include a mass term and interaction terms away from the horizon 
because it is difficult to construct the conformal field theory with a mass or interactions in 4-dimensions.
By contrast, we can consider the 4-dimensional theory including a mass term and interaction terms away from the horizon as in (\ref{KNact01}).
Secondly, we can derive not only the Hawking temperature but also the Hawking black body spectrum.
This means that the black hole behaves not merely as the thermal body but as the black body.
This observation agrees with Hawking's original derivation.

Finally we comment on the black hole entropy.
We defined quantum states as in (\ref{sta}) and obtained the reduced density matrix as in (\ref{red}).
We can evaluate the entropy for these states by using the von Neumann entropy formula.
However, we must mention that it is not the entropy of the black hole itself but rather the entropy of the boson field.
Our method does not allow us to derive the entropy for a black hole with a finite temperature by counting the number of quantum states associated with the black hole.
Therefore, we derived the black hole entropy from thermodynamic considerations as in (\ref{ent01}) following previous works.
The derivation of the black hole entropy by counting the number of black hole quantum states remains as one of future problems.


\section*{Acknowledgements}

The present work was initiated when I was visiting S. N. Bose National Centre for Basic Sciences, Kolkata.
I am happy to thank Prof. R. Banerjee and Dr. B. R. Majhi for patiently explaining me the basic idea of
their tunneling mechanism. 
I thank Dr. S. Kulkarni and other members of Bose Center for thier hospitality extended to me. 
I also  thank Prof. K. Fujikawa for helpful discussions and for a careful reading of the manuscript. 
Helpful comments on the manuscript by Prof. Banejee are gratefully acknowledged. 
The present work is supported by the strategic project for academic research of Nihon University.

%
\appendix
\section{Kruskal-like coordinate}

In this appendix, we show how to obtain the Kruskal-like coordinates (\ref{Kru01}) and (\ref{Kru02}) from the $(t-r)$-coordinates which appears in the 2-dimensional spherically-symmetric metric (\ref{kerr5}) by performing several coordinate transformations.
As a preliminary analysis, we would like to summarize the Schwarzschild metric and Kruskal-Szekeres coordinates briefly \cite{Kru,Sze}.
The well-known Schwarzschild metric is given by
\begin{align}
ds^2=-\left( 1-\frac{2M}{r} \right) dt^2 + \frac{1}{1-\frac{2M}{r}}dr^2+ r^2(d\theta^2+\sin^2 \theta d\varphi^2).
\label{sch01}
\end{align}
It follows from the expression (\ref{sch01}) that there are two singularities $r=0$ and $r=2M$ in the Schwarzschild metric.
A singularity at $r=0$ is the curvature singularity which cannot be removed while the other at $r=2M$ is a fictitious singularity
arising merely from an improper choice of coordinates.
Therefore the singularity at $r=2M$ can be removed by using appropriate coordinates.
Kruskal and Szekeres use a dimensionless time coordinate $T$ and a dimensionless radial coordinate $R$
related to the Schwarzschild $t$ and $r$ by
\begin{align}
\left. \begin{array}{ccc}
T&=\left( \frac{r}{2M} -1 \right)^{\frac{1}{2}} e^{\frac{r}{4M}} \sinh \left( \frac{t}{4M}\right)\\
R&=\left( \frac{r}{2M} -1 \right)^{\frac{1}{2}} e^{\frac{r}{4M}} \cosh \left( \frac{t}{4M}\right)
\end{array}
\right\} \quad {\rm when} \quad r >2M,
\label{apkru03}\\
\left. \begin{array}{ccc}
T&=\left( 1-\frac{r}{2M}  \right)^{\frac{1}{2}} e^{\frac{r}{4M}} \cosh \left( \frac{t}{4M}\right)\\
R&=\left( 1-\frac{r}{2M}  \right)^{\frac{1}{2}} e^{\frac{r}{4M}} \sinh \left( \frac{t}{4M}\right)
\end{array}
\right\} \quad {\rm when} \quad r <2M.
\label{apkru04}
\end{align}
By making this change of coordinates in the Schwarzschild metric (\ref{sch01}), we obtain
\begin{align}
ds^2=\frac{32M^3}{r} e^{-\frac{r}{2M}} (-dT^2+dX^2)+ r^2(d\theta^2+\sin^2 \theta d\varphi^2).
\label{sch02}
\end{align}
It follows from the expression (\ref{sch02}) that there is only the curvature singularity at $r=0$, i.e., 
the singularity at $r=2M$ does not appear. 
The coordinates defined as in (\ref{apkru03}) and (\ref{apkru04}), are called Kruskal coordinates.
Similarly, we can also perform related discussions for the spherically-symmetric metric such as the Reissner-Nordstr\"om metric,
and the defined coordinates are then called "the Kruskal-like coordinates" in common parlance.

In the body of the paper, we showed that the 4-dimensional Kerr-Newman metric becomes the 2-dimensional spherically-symmetric 
metric (\ref{kerr5}) by using a technique of the dimensional reduction near the horizon.
In what follows, we derive the Kruskal-like coordinates for this reduced metric 
following the derivation of the Kruskal coordinates in a standard textbook.
In this derivation, we apply a series of coordinate transformations.
Note that all the variables which appear in our transformations are defined to be real numbers.

Now the metric is given by
\begin{align}
ds^2=-F(r)dt^2+\frac{1}{F(r)}dr^2,
\label{ds01}
\end{align}
where $F(r)$ is
\begin{align}
F(r)=\frac{(r-r_+)(r-r_-)}{r^2+a^2},
\label{apkru02}
\end{align}
and $r_{+(-)}$ is the outer (inner) horizon given by (\ref{KN05}).

As a first step of coordinate transformations, we use the tortoise coordinate defined by \cite{wh}
\begin{align}
dr_*\equiv \frac{1}{F(r)}dr.
\label{r*def01}
\end{align}
The metric (\ref{ds01}) is then written by
\begin{align}
ds^2=-F(r)(dt-dr_*)(dt+dr_*).
\label{ds02}
\end{align}
By integrating (\ref{r*def01}) over $r$ from 0 to r, we obtain
\begin{align}
r_*=r+\frac{1}{2K_+}\ln \frac{|r-r_+|}{r_+}+\frac{1}{2K_-}\ln\frac{|r-r_-|}{r_-}+C,
\label{r*def03}
\end{align}
where $K_{+(-)}$ is the surface gravity on the outer (inner) horizon and $C$ is generally a pure imaginary integral constant which appears in the analytic continuation.
However, as already stated, we would like to treat the case where all the variables (or parameters) are defined in the range of real numbers.
Thus we need to consider the three cases $r>r_+$, $r_+<r<r_-$ and $r_-<r<0$ with respect to the range of $r$.
Actually, we have only to consider the two cases $r>r_+$ and $r_+<r<r_-$ because of the consideration near the outer horizon.
When $r>r_+$, the relation (\ref{r*def03}) becomes
\begin{align}
r_*=r+\frac{1}{2K_+}\ln \frac{r-r_+}{r_+}+\frac{1}{2K_-}\ln\frac{r-r_-}{r_-}.
\label{r*def04}
\end{align}
As the second step we use the retarded time $u$ and the advanced time $v$ defined by
\begin{align}
\left.
\begin{array}{ccc}
u&\equiv t-r_*=t-r+\frac{1}{2K_+}\ln \frac{r-r_+}{r_+}+\frac{1}{2K_-}\ln\frac{r-r_-}{r_-}\\
v&\equiv t+r_*=t+r+\frac{1}{2K_+}\ln \frac{r-r_+}{r_+}+\frac{1}{2K_-}\ln\frac{r-r_-}{r_-}
\end{array}\right\} \quad {\rm when} \quad r>r_+.
\end{align}
The metric (\ref{ds02}) is then written as
\begin{align}
ds^2=-F(r)dudv.
\label{ds03}
\end{align}
As the third step we use the following coordinate transformations $U$ and $V$ defined by
\begin{align}
\left.
\begin{array}{ccc}
&U\equiv -e^{-K_+ u}=-\left( \frac{r-r_+}{r_+} \right)^{\frac{1}{2}} \left( \frac{r-r_-}{r_-}\right)^{\frac{K_+}{2K_-}}e^{K_+ r} e^{-K_+t}\\
&V\equiv e^{K_+ v}=\left( \frac{r-r_+}{r_+} \right)^{\frac{1}{2}} \left( \frac{r-r_-}{r_-}\right)^{\frac{K_+}{2K_-}}e^{K_+ r} e^{K_+t}\qquad \quad
\end{array}\right\} \quad {\rm when} \quad r>r_+.
\end{align}
The metric (\ref{ds03}) is then written as
\begin{align}
ds^2=-\frac{r_+r_-}{K_+^2} \frac{e^{-2K_+ r}}{r^2+a^2} \left( \frac{r_-}{r-r_-}\right)^{\frac{K_+}{K_-} -1} dUdV.
\label{ds04}
\end{align}
As the final step we use the following coordinate transformations $T$, $R$ defined by
\begin{align}
\left.
\begin{array}{ccc}
T \equiv \frac{1}{2}(V+U)=\left( \frac{r-r_+}{r_+} \right)^{\frac{1}{2}} \left( \frac{r-r_-}{r_-}\right)^{\frac{K_+}{2K_-}}e^{K_+ r} \sinh (K_+ t)\\
R \equiv \frac{1}{2}(V-U)=\left( \frac{r-r_+}{r_+} \right)^{\frac{1}{2}} \left( \frac{r-r_-}{r_-}\right)^{\frac{K_+}{2K_-}}e^{K_+ r} \cosh (K_+ t)
\end{array}\right\} \quad {\rm when} \quad r>r_+.
\label{apkru01}
\end{align}
The metric (\ref{ds04}) is then written as
\begin{align}
ds^2=\frac{r_+r_-}{K_+^2} \frac{e^{-2K_+ r}}{r^2+a^2} \left( \frac{r_-}{r-r_-}\right)^{\frac{K_+}{K_-} -1} (-dT^2+dR^2).
\label{ds05}
\end{align}

Similarly, we consider the case of $r_+>r>r_-$.
When $r_+>r>r_-$, the relation (\ref{r*def03}) becomes
\begin{align}
r_*=r+\frac{1}{2K_+}\ln \frac{r_+-r}{r_+}+\frac{1}{2K_-}\ln\frac{r-r_-}{r_-}.
\label{r*def05}
\end{align}
As for the remaining coordinate transformations, we use the following ones
\begin{align}
\left.
\begin{array}{ccc}
u&\equiv t-r_*=t-r+\frac{1}{2K_+}\ln \frac{r_+-r}{r_+}+\frac{1}{2K_-}\ln\frac{r-r_-}{r_-}\\
v&\equiv t+r_*=t+r+\frac{1}{2K_+}\ln \frac{r_+-r}{r_+}+\frac{1}{2K_-}\ln\frac{r-r_-}{r_-}
\end{array}\right\} \quad {\rm when} \quad r_+>r>r_-,
\end{align}
\begin{align}
\left.
\begin{array}{ccc}
&U\equiv e^{-K_+ u}=\left( \frac{r_+-r}{r_+} \right)^{\frac{1}{2}} \left( \frac{r-r_-}{r_-}\right)^{\frac{K_+}{2K_-}}e^{K_+ r} e^{-K_+t}\\
&V\equiv e^{K_+ v}=\left( \frac{r_+-r}{r_+} \right)^{\frac{1}{2}} \left( \frac{r-r_-}{r_-}\right)^{\frac{K_+}{2K_-}}e^{K_+ r} e^{K_+t}\quad ~
\end{array}\right\} \quad {\rm when} \quad r_+>r>r_-,
\end{align}
\begin{align}
\left.
\begin{array}{ccc}
T \equiv \frac{1}{2}(V+U)=\left( \frac{r_+-r}{r_+} \right)^{\frac{1}{2}} \left( \frac{r-r_-}{r_-}\right)^{\frac{K_+}{2K_-}}e^{K_+ r} \cosh (K_+ t)\\
R \equiv \frac{1}{2}(V-U)=\left( \frac{r_+-r}{r_+} \right)^{\frac{1}{2}} \left( \frac{r-r_-}{r_-}\right)^{\frac{K_+}{2K_-}}e^{K_+ r} \sinh (K_+ t)
\end{array}\right\} \quad {\rm when} \quad r_+>r>r_-.
\label{apkru02}
\end{align}
Of course, by performing these coordinate transformations, the corresponding metrics (\ref{ds03}), (\ref{ds04}) and (\ref{ds05}) 
do not change.
The two sets of defined coordinates, both (\ref{apkru01}) and (\ref{apkru02}), are in fact the Kruskal-like coordinates.
In the Schwarzshild case ($a=Q=0$), (\ref{apkru01}) and (\ref{apkru02}) respectively agree with (\ref{apkru03}) and (\ref{apkru04}).
Finally, we obtain (\ref{Kru01}) and (\ref{Kru02}) by rewriting the expressions written in terms of $r$ to the ones in terms of $r_*$ in the formulas (\ref{apkru01}) and (\ref{apkru02}).

\end{document}